\documentclass[aps,prd,twocolumn,groupedaddress,showpacs]{revtex4}
\usepackage{graphicx}% Include figure files
\usepackage{dcolumn}% Align table columns on decimal point
\usepackage{bm}% bold math

\begin{document}

\title{
 Superluminal pions in a hadronic fluid}

\author{Neven Bili\'c}
\email{bilic@thphys.irb.hr}
\author{Hrvoje Nikoli\'c}
\affiliation{
Rudjer Bo\v{s}kovi\'{c} Institute,
P.O. Box 180, 10002 Zagreb, Croatia
}

\date{\today}

\begin{abstract}
We study the propagation of pions
at finite temperature and finite chemical potential
in the framework of the linear
sigma model with 2 quark flavors and
$N_c$ colors.
The velocity of massless pions in general
differs from that of light.
 One-loop calculations show that
 in the chiral symmetry broken phase pions,
 under certain conditions,
 propagate faster than light.
\end{abstract}


\maketitle

\section{Introduction}

The linear sigma model,
originally proposed as a  model for
strong nuclear interactions
\cite{gel},
today serves as an effective
model for the low-energy (low-temperature)
phase of quantum chromodynamics.
The model exhibits spontaneous breaking
of chiral symmetry
 and restoration
at finite temperature.
In the chiral symmetry broken phase at zero temperature
pions, being massless,
propagate with the velocity of light.
At finite temperature
one would expect
chiral pions
to propagate slower than light,
owing to medium effects.
Indeed, this expectation has been confirmed by
one-loop
calculations in the linear sigma model with
no fermions \cite{pis2,son1,son2,aya}.

In this paper we discuss
the propagation of pions at nonzero temperature and
nonzero finite baryon density in a model with two quark flavors.
It turns out that  pions in the presence of fermions
become superluminal
in a certain range of temperature and baryon chemical
potential.

A superluminal propagation
has recently been studied in the context of
the Casimir effect in quantum electrodynamics.
Scharnhorst has demonstrated
\cite{sch} that, when vacuum fluctuations
of the electromagnetic field obey periodic boundary conditions in one
spatial dimension (corresponding to parallel Casimir plates),
then the two-loop corrections to the
photon polarization tensor lead to a superluminal propagation of
photons.
A similarity between the effects of Casimir plates
and that of finite temperature on the propagation of photons
has been discussed \cite{bart,lat}, and more general conditions
that lead to a superluminal propagation of photons have
also
been identified.

A similar effect has been found for photons
interacting with fermions
\cite{fer}.
It has been shown that the transverse
photons coupled to fermions obeying periodic boundary
conditions
are  faster than light
when propagating perpendicularly
and are slower than light
when  propagating
parallelly to
the compactified dimension.
Since
our fermions obey the usual antiperiodic boundary conditions
in  the compact temporal dimension,
one would  expect no superluminal effects.
However, we demonstrate that in a certain region of temperature
and chemical potential
below the symmetry restoration point,
 massless pions propagate faster than light.
Moreover,
if one of the spatial dimensions is compactified,
pions will propagate superluminally or subluminally,
depending on the size of the compact dimension
and on the boundary conditions.

We organize the paper as follows.
In Sec.\ \ref{eff} we describe
the model.
In Sec.\ \ref{velocity}
 we calculate the dependence of the pion velocity on temperature
 and chemical potential.
 We present
the results and discussion in
Sec.\ \ref{results}.
 In the concluding section,
Sec.\ \ref{concl}, we summarize our results.

\section{Effective Lagrangian} \label{eff}

Consider the linear sigma model at finite temperature
and finite baryon density.
The thermal bath provides a medium which may
be assumed to have a homogeneous velocity field.
The dynamics of mesons in such a medium is described by
an effective chirally symmetric Lagrangian of the form
\begin{eqnarray}
\label{eq100}
{\cal{L}}
       \! & \! = \! & \!
 \frac{1}{2}(a\, \eta^{\mu\nu}
 +b\, u^{\mu}u^{\nu})(\partial_{\mu} \sigma
 \partial_{\nu} \sigma
+\partial_{\mu}
\mbox{\boldmath $\pi$}
\partial_{\nu}
\mbox{\boldmath $\pi$})
  \nonumber \\
  & &
 - \frac{m_0^{2}}{2} (\sigma^{2} +
\mbox{\boldmath$\pi$}^{2})
-\frac{\lambda}{4} (\sigma^{2} +
\mbox{\boldmath$\pi$}^{2})^{2}
 \nonumber  \\
         &   &
+\bar{\psi} (c\, i \gamma^{\mu} \partial_{\mu} +
\mu u^{\mu}\gamma_{\mu} +
g (\sigma + i \mbox{\boldmath $\tau \pi$} \gamma_5))\psi
\, ,
 \nonumber  \\
         &   &
\end{eqnarray}
where $u_{\mu}$ is the velocity of the fluid,
and $\eta_{\mu\nu}$ is the flat Lorentzian metric tensor.
In the original sigma model \cite{gel} the fermion field
was a nucleon.
We consider
 fermions to be
constituent quarks \cite{con,goc,baie,bil1,nem,cal} with
additional $N_c$ degrees of freedom,
``colors'', from the SU($N_{c}$) local gauge group of
an underlying gauge theory (QCD).
The parameters
 $a$, $b$,
 and $c$ depend on the temperature $T$,
 the chemical potential $\mu$,
 and the parameters of the model $m_0$, $\lambda$,
 and $g$,
 and may be calculated
in perturbation theory.
At $T=\mu=0$ the medium is absent and $
a=c=1$ and $b=0$.

 If $m_0^{2} < 0$,  chiral
symmetry will be spontaneously broken.
At the classical level, the $\sigma$ field develops a
nonvanishing expectation value
$\langle \sigma \rangle =f_{\pi}$.
At nonzero temperature the expectation value
$\langle \sigma \rangle$  depends on the temperature
and vanishes at the chiral transition point.
 Redefining the fields as
\begin{equation}\label{eq9}
(\sigma,\mbox{\boldmath$\pi$})
\rightarrow
(\sigma,\mbox{\boldmath$\pi$})+
 (\sigma'(x),\mbox{\boldmath$\pi$}'(x)) ,
\end{equation}
where {\boldmath$\pi'$}
and $\sigma'$
are quantum fluctuations around the
constant values $\mbox{\boldmath$\pi$}=0$ and $\sigma
=\langle \sigma \rangle$,
respectively,
we obtain
 the effective Lagrangian
 in which
chiral symmetry is spontaneously broken:
\begin{widetext}
\begin{eqnarray}\label{eq5}
{\cal{L}'}
       \! & \! = \! & \!
 \frac{1}{2}(a\, \eta^{\mu\nu}
 +b\, u^{\mu}u^{\nu})
\partial_{\mu}
\mbox{\boldmath $\pi$}'
\partial_{\nu}
\mbox{\boldmath $\pi$}'
 +\frac{1}{2}(\tilde{a}\, \eta^{\mu\nu}
 +\tilde{b}\, u^{\mu}u^{\nu})
 \partial_{\mu} \sigma'
 \partial_{\nu} \sigma'
- \frac{m_{\sigma}^{2}}{2} \sigma'^{2}
- \frac{m_{\pi}^{2}}{2}
\mbox{\boldmath$\pi$}'^{2}
-g' \sigma' (\sigma'^2 +
\mbox{\boldmath $\pi$}'^{2})
\nonumber  \\
         &   &
- {\lambda\over 4}
 (\sigma'^{2} +\mbox{\boldmath$\pi$}'^{2})^2
+\bar{\psi} (c \, i \gamma^{\mu} \partial_{\mu}
+m_F +\mu u^{\mu}\gamma_{\mu} +
g (\sigma' + i \mbox{\boldmath $\tau \pi$}' \gamma_5))\psi
\; .
\end{eqnarray}
\end{widetext}
At  temperatures and chemical potentials below the
chiral transition line
the trilinear coupling and the masses are given by
\begin{eqnarray}\label{eq43}
m_{\sigma}^{2} & = &
 2 \lambda \sigma^2  \,  ,  \; \;\;\;
  m_{F} =g \sigma \, ,
  \nonumber \\
m_{\pi}^{2} & = & 0 \, , \;\;\;\;
g' =  \lambda \sigma  \, .
\end{eqnarray}
in agreement with the Goldstone theorem.

The boson kinetic part in Eq.\ (\ref{eq5}) is split in two
   terms since the pion and sigma self-energies get different
   finite temperature and chemical potential contributions
   in the chiral-symmetry broken phase.
   Hence, in general, $a\neq \tilde{a}$ and $b\neq \tilde{b}$,
   at temperatures and chemical
   potentials below the chiral transition line.
   The constant $c$ is related to the
   finite temperature and chemical potential
   contributions to the fermion self-energy
   and is irrelevant to the calculation of the pion velocity.
   As we shall shortly see, the only quantities
   relevant to the calculation of the pion velocity
   are the constants $a$ and $b$ that enter the pion kinetic term
   in Eq.\ (\ref{eq5}).

The temperature dependence of the chiral condensate $\sigma$
is obtained by
minimizing the thermodynamic potential
$\Omega=-(T/V) \ln Z$
with respect to
 $\sigma$
at fixed inverse temperature $\beta$.
At one-loop order,
the extremum condition
reads
\cite{bil1}
\begin{eqnarray}\label{eq032}
\sigma^{2} \!&\! = \!&\!
f_{\pi}^{2} - {8 g^{2}\over \lambda}     N_{c}
\: \int {d^{3} p\over (2 \pi)^{3}}
\: {1\over 2\omega_F} \;  n_F (\omega_F)
 \nonumber  \\
 &   &
- 3 \! \int\!  {d^{3} p\over (2 \pi)^{3}}
 {1\over \omega_{\sigma}}  n_{B} (\omega_{\sigma})
- 3 \! \int\! {d^{3} p\over (2 \pi)^{3}}
 {1\over \omega_{\pi}}  n_{B} (\omega_{\pi}) ,
  \nonumber \\
 & &
\end{eqnarray}
where
\begin{eqnarray}\label{eq28}
\omega_{\sigma}
\!&\! = \!&\! (\mbox{\boldmath $p$}^2+m_{\sigma}^{2})^{1/2}
\, , \;\;\;\;
\omega_{\pi}
=|\mbox{\boldmath $p$}|
\, , \;\;\;\;
\nonumber \\
\omega_F \!&\! = \!&\!
(\mbox{\boldmath $p$}^2+m_F^{2})^{1/2},
\end{eqnarray}
\begin{equation}\label{eq29}
n_{F}(\omega) = {1\over e^{\beta(\omega - \mu)} + 1} +
{1\over e^{\beta(\omega + \mu)} + 1}  \, ,
\end{equation}
\begin{equation}\label{eq30}
n_{B}(\omega) = {1\over e^{\beta \omega} - 1}     \, .
\end{equation}
The right-hand side of (\ref{eq032}) depends on
$\sigma$ through the masses
$m_{\sigma}$ and $m_F$ given by (\ref{eq43}).
Solutions to  (\ref{eq032})
 are implicit functions of $T$ and $\mu$.
These equations have been derived from the
thermodynamic  potential
in which the loop corrections
have been neglected \cite{bil1}.
This approximation corresponds to the
leading order in the
$1/N$ expansion, where $N$ is the
number of scalar fields \cite{mey}.
In our case, $N=4$.

\section{Pion velocity}
\label{velocity}
The propagation of pions is governed by the equation of motion
\begin{equation}
\partial_{\mu}
\left[
\,
( a\, \eta^{\mu\nu}+ b\,
u^{\mu}
u^{\nu})\right] \partial_{\nu}\mbox{\boldmath{$\pi$}}
+V(\sigma,
\mbox{\boldmath{$\pi$}},\psi)
\mbox{\boldmath{$\pi$}}=0 \, ,
\label{eq013}
\end{equation}
where $V$ is the interaction potential
the form of which is irrelevant  to our
consideration.

In  the simplest case of
   a homogeneous flow,
  Eq.\ (\ref{eq013}) reduces to
  the  wave equation
\begin{equation}
(\partial_t^2 -
v^2
\Delta +\frac{v^2}{a}V)
\mbox{\boldmath{$\pi$}}=0,
\label{eq014}
\end{equation}
where the quantity $v$ is the pion velocity
given by
\begin{equation}
v^2=\left(1+\frac{b}{a}\right)^{-1}  .
\label{eq015}
\end{equation}
As we shall shortly demonstrate, the constants $a$ and $b$
may be derived from the finite-temperature perturbation
expansion of the pion self-energy.

The pion velocity in a sigma model at finite temperature
has been calculated at one-loop level by Pisarski and Tytgat
in the low-temperature approximation
\cite{pis2} and by Son and Stephanov for temperatures
close to the chiral transition point \cite{son1,son2}.
It has been found that the pion velocity vanishes
as one approaches the critical temperature.
Here we analyze the whole range of temperatures in the
chiral symmetry broken phase.

Consider the pion
 self-energy  $\Sigma(q,T)$
in the limit when the external momentum
$q$ approaches 0.
The renormalized inverse pion propagator
may, in the limit $q\rightarrow 0$, be expressed in the form
\begin{eqnarray}
 Z_{\pi}\Delta^{-1} \! & \! = \! & \!
  q^{\mu}q_{\mu}-
 \frac{1}{2!}
 q^{\mu}q^{\nu}\frac{\partial}{\partial q^{\mu}}
 \frac{\partial}{\partial q^{\nu}}
\left. (\Sigma(q,T)
-  \Sigma(q,0)) \right|_{q=0}
 \nonumber \\
 & & +\dots   \, ,
  \label{eq201}
\end{eqnarray}
where the ellipsis denotes the terms of higher order in
$q^{\mu}$.
The $q^{\mu}$ independent term of the self-energy
absorbs in the renormalized pion mass,
which is equal to zero in the chiral symmetry broken phase.
The subtracted T=0 term  has been
absorbed in the wave function renormalization factor $Z_{\pi}$.
By comparing this equation with
the inverse pion propagator derived directly from
the effective Lagrangian (\ref{eq5})
\begin{equation}
 \Delta^{-1}=(a+b)q_0^2
 -a \mbox{\boldmath $q$}^2,
  \label{eq202}
\end{equation}
we can express
 the parameters $a$ and $b$,
 and hence the pion
velocity, in terms of the second derivatives of
$\Sigma(q,T)$ evaluated at $q^{\mu}=0$.
At one-loop level the diagrams that
give a nontrivial $q$-dependence of $\Sigma$ are the bubble
diagrams. Subtracting the T=0 term
 one finds
\begin{equation}
\Sigma(q)
  \equiv
\Sigma(q,T)
-  \Sigma(q,0)
=\Sigma_B
+  \Sigma_F
  \label{eq203}
\end{equation}
with the contribution of the bosonic and fermionic loops
given by
\begin{widetext}
\begin{eqnarray}
\Sigma_B(q)
  \!&\! = \!&\!
  -4g'^2 \int\!
\frac{d^3p}{(2\pi)^3}
\frac{1}{2\omega_{\pi}
2\omega_{\sigma,q}}
%\nonumber\\
%  \!&\!\!&\! \times
  \left\{ [n_B(\omega_{\pi})+
n_B(\omega_{\sigma,q})]
\left(\frac{1}{\omega_{\sigma,q}+
\omega_{\pi}-q_0}
  + \frac{1}{\omega_{\sigma,q}+
\omega_{\pi}+q_0}\right)\right. \nonumber\\
  \!&\!\!&\!
   + \left. [n_B(\omega_{\pi})-
n_B(\omega_{\sigma,q})]\left(
  \frac{1}{\omega_{\sigma,q}-
\omega_{\pi}-q_0} +
  \frac{1}{\omega_{\sigma,q}-
\omega_{\pi}+ q_0}\right)\right\},
  \label{eq213}
\end{eqnarray}
\begin{eqnarray}
\Sigma_F(q)
  \!&\! = \!&\!
-8N_cg^2 \int\!\frac{d^3p}{(2\pi)^3} \frac{1}{2\omega_F2\omega_{F,q}}
%\nonumber\\
%\!&\!\!&\! \times
\left\{ \frac{-q_0\omega_{F,q}+\mbox{\boldmath $q$}
(\mbox{\boldmath $p$}+\mbox{\boldmath $q$})}{e^{\beta(\omega_{F,q}-\mu)
}+1}
\left( \frac{1}{\omega_F+\omega_{F,q}-q_0}
+\frac{1}{\omega_F-\omega_{F,q}+q_0} \right)
\right. \nonumber\\
\!&\!\!&\!
+\frac{-q_0^2-q_0\omega_{F}+\mbox{\boldmath $q$}(\mbox{\boldmath $p$}
+\mbox{\boldmath $q$})}{e^{\beta(\omega_{F}-\mu)}+1}
\left( \frac{1}{\omega_{F,q}+\omega_F+q_0}
+\frac{1}{\omega_{F,q}-\omega_F-q_0} \right)
\nonumber\\
\!&\!\!&\!
+\frac{q_0\omega_{F,q}+\mbox{\boldmath $q$}(\mbox{\boldmath $p$}
+\mbox{\boldmath $q$})}{e^{\beta(\omega_{F,q}+\mu)}+1}
\left( \frac{1}{\omega_F+\omega_{F,q}+q_0}
+\frac{1}{\omega_F-\omega_{F,q}-q_0} \right)
\nonumber\\
\!&\!\!&\! \left.
+\frac{-q_0^2+q_0\omega_{F}+\mbox{\boldmath $q$}(\mbox{\boldmath $p$}
+\mbox{\boldmath $q$})}{e^{\beta(\omega_{F}+\mu)}+1}
\left( \frac{1}{\omega_{F,q}+\omega_F-q_0}
+\frac{1}{\omega_{F,q}-\omega_F+q_0} \right)
\right\}  \, ,
  \label{eq223}
\end{eqnarray}
where
$\omega_{\sigma,q}=
[(\mbox{\boldmath $p$}-
\mbox{\boldmath $q$})^2
+m_\sigma^2]^{1/2}$,
$\omega_{F,q}=[(\mbox{\boldmath $p$}+
 \mbox{\boldmath $q$})^2
+m_F^2]^{1/2}$.
A straightforward evaluation of the second derivatives of
$\Sigma(q)$ at $q_{\mu}=0$ yields
\begin{equation}
a=1+a_B+a_F  \, ,
\label{eq301}
\end{equation}
\begin{equation}
b=b_B+b_F      \, ,
\label{eq302}
\end{equation}
with
\begin{eqnarray}
a_B  =
   \frac{16 g'^2}{m_{\sigma}^4} \int\!
\frac{d^3p}{(2\pi)^3}
  \left[ \frac{n_B(\omega_{\pi})}{4\omega_{\pi}}+
\frac{n_B(\omega_{\sigma})
}{4\omega_{\sigma}}
 - \frac{1}{3}
      \frac{\omega_{\pi}^2}{m_{\sigma}^2}
   \left(
   \frac{n_B(\omega_{\pi})}{\omega_{\pi}} -
\frac{n_B(\omega_{\sigma})
}{\omega_{\sigma}}
\right)\right] \, ,
\label{eq204}
%\end{equation}
\end{eqnarray}
%\begin{equation}
\begin{eqnarray}
%b =
b_B =
   \frac{16g'^2}{m_{\sigma}^4} \int\!
\frac{d^3p}{(2\pi)^3}
  \left[
  \frac{\omega_{\pi} n_B(\omega_{\pi})
  }{m_{\sigma}^2}-
\frac{\omega_{\sigma}
n_B(\omega_{\sigma})
  }{m_{\sigma}^2}
 + \frac{1}{3}
      \frac{\omega_{\pi}^2}{m_{\sigma}^2}
   \left(
   \frac{n_B(\omega_{\pi})}{\omega_{\pi}} -
\frac{n_B(\omega_{\sigma})
}{\omega_{\sigma}}
\right)\right]  \, ,
 \label{eq205}
%\end{equation}
\end{eqnarray}
\end{widetext}
\begin{eqnarray}
a_F =
N_cg^2\int\!\frac{d^3p}{(2\pi)^3}
\frac{n_F(\omega_F)}{p^2\omega_F}\,,
 \label{eq304}
\end{eqnarray}
\begin{eqnarray}
b_F =
-N_cg^2\int\!\frac{d^3p}{(2\pi)^3}
\frac{m_F^2}{p^2}
\frac{n_F(\omega_F)}{\omega_F^3}\, .
 \label{eq305}
%\end{equation}
\end{eqnarray}
%\end{widetext}

\noindent
The subscripts $B$ and $F$ denote
the contributions of
the boson and fermion loops,
respectively.
The sign of the fermion contribution in the last equation
is crucial.
The pion velocity $v$ given by Eq.\ (\ref{eq015}) will
become larger than unity when $b<0$, i.e., when the negative fermionic
part exceeds the positive bosonic part.

The results (\ref{eq301}) and (\ref{eq302})
with
(\ref{eq204})-(\ref{eq305})
can easily be extended to include
the sigma model at $T=0$ and $\mu=0$ in
3+1 dimensional space-time with one
spatial dimension, e.g., in the $z$-direction,
compactified
to the size
 $L\equiv\beta$.
In this case, the effective Lagrangian
may be written in the form of Eq.\ (\ref{eq100})
with $b$ replaced by $-b$
and the velocity  $u_{\mu}$ replaced by
a spacelike vector $n_{\mu}$
parallel to the compactified dimension and normalized as
$n^{\mu} n_{\mu}=-1$. The inverse propagator is now given by
\begin{equation}
\Delta^{-1}=a (q_0^2-q_x^2-q_y^2)-(a+b) q_z^2
\end{equation}
and hence the velocity in the compact direction
precisely equals the inverse of $v$ given by (\ref{eq015}).
Clearly, the velocity in the subspace orthogonal to the compact
dimension is equal to the velocity of light.
Recall that the extension of field theory to nonzero
   temperature is equivalent to a compactification of
   the time coordinate. Now, the compactified spatial coordinate
   and the time coordinate exchange their roles.
   However, since there is no time-ordering restriction
   in spatial coordinates, fermions may be chosen to be
   periodic or antiperiodic in the compact dimension.
   Hence, in the case of antiperiodic fermions,
   all the calculations of the self-energy are exactly the same
as in the finite temperature case with $T=1/L$.
   The case of periodic fermions is similar, but one
 has to take into account that the replacement of the antiperiodic
   boundary condition at $\mu=0$ by the periodic one
   modifies the $\beta$-dependent part of the loop sum over 4-momenta
   so that the Fermi-Dirac function $1/(1+e^{\beta\omega})$
   in $n_F$ is replaced by $1/(1-e^{\beta\omega})$.
   In this way the pion velocity parallel to the
   compact direction may be written as
\begin{equation}
v_{||}^2=1+\frac{b'}{a'}\, ,
\label{eq306}
\end{equation}
where $a'=a$ and $b'=b$ for antiperiodic and
\begin{equation}
a'=1+a_B-\bar{a}_F  \, ,
\label{eq401}
\end{equation}
\begin{equation}
b'=b_B-\bar{b}_F      \, ,
\label{eq402}
\end{equation}
for periodic fermions, with
\begin{eqnarray}
\bar{a}_F =
2 N_cg^2\int\!\frac{d^3p}{(2\pi)^3}
\frac{n_B(\omega_F)}{p^2\omega_F}\,,
 \label{eq404}
\end{eqnarray}
\begin{eqnarray}
\bar{b}_F =
-2 N_cg^2\int\!\frac{d^3p}{(2\pi)^3}
\frac{m_F^2}{p^2}
\frac{n_B(\omega_F)}{\omega_F^3}\, .
 \label{eq405}
 \end{eqnarray}
Note that in the case of periodic fermions
Eq.\ (\ref{eq032}) also modifies, because
$n_F(\omega_F)$ must be replaced by
$-2 n_B(\omega_F)$. Depending on the choice of the coupling
constants $g$ and $\lambda$, the symmetry breaking and restoration
pattern may be quite different.
In particular, the temperature $T_c'$ at which
   sigma vanishes  and which approximates the critical
   temperature \cite{bil1}, is now given by
\begin{equation}
 \left(\frac{T_c'}{f_{\pi}}\right)^2=
\frac{6 \lambda}{-2g^2 N_c +3 \lambda} \, .
 \label{eq505}
\end{equation}
   Clearly, for $\lambda < (2/3) g^2 N_c$
   there is no restoration of chiral
   symmetry.

\begin{figure*}[t]
\includegraphics{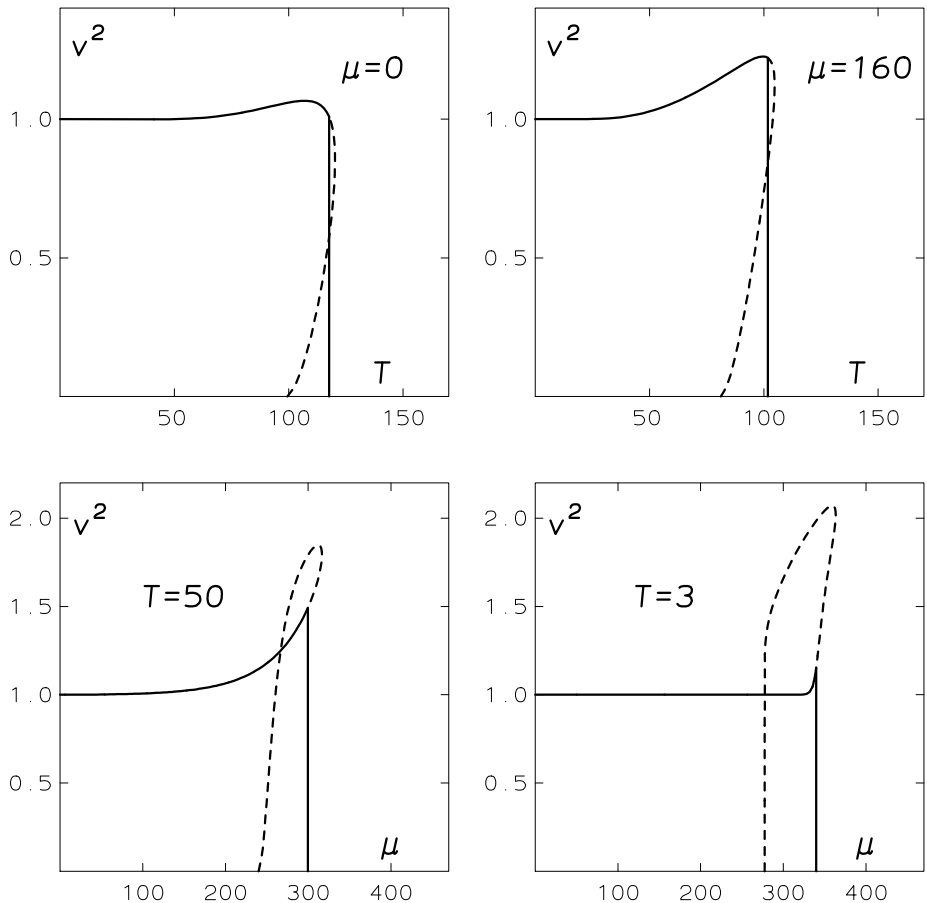}
\caption{
Pion velocity squared as a function of temperature and
chemical potential.
The dashed line corresponds to a thermodynamically unstable
solution. $T$ and $\mu$ are in MeV.}
\label{fig1}
\end{figure*}

\section{Results and discussion}\label{results}

   In order to proceed with
a quantitative analysis, we have to choose the input parameters
from  particle physics phenomenology.
For the constituent quark mass we take
$m_F=340$ MeV.
The coupling
$g$  is then fixed by (\ref{eq43})
$g=m_F/f_{\pi}$.
The  Particle Data Group
gives a rather wide range
400-1200 MeV \cite{part} of the sigma meson masses.
We shall shortly see that the
analysis of the $T=0$, $\mu\neq 0$ case
further restricts this available range.

 At $T=0$ the extremum condition (\ref{eq032}) for
nonnegative $\mu$ and $N_c=3$ reads
\begin{widetext}
\begin{equation}
\sigma^2 =f_{\pi}^2 -\frac{3g^2}{\lambda\pi^2}\:\vartheta (\mu -
 g\sigma)\left[ \mu\sqrt{\mu^2 -g^2\sigma^2}+g^2\sigma^2\ln
 \frac{g\sigma}{\mu +\sqrt{\mu^2 -g^2\sigma^2}} \right] \, ,
  \label{eq206}
\end{equation}
which determines the extremum of the thermodynamic potential
%\begin{eqnarray}
\begin{equation}
\Omega(\sigma,\mu)
%\!&\! = \!&\!
=
\frac{\lambda}{4}\sigma^4
-\frac{\lambda}{2}f_{\pi}^2\sigma^2
%\nonumber\\
%\!&\!\!&\!
-\frac{\vartheta (\mu -g\sigma)}{(2\pi)^2}
\left[ \mu\sqrt{\mu^2-g^2\sigma^2}(2\mu^2-5g^2\sigma^2)
+3g^4\sigma^4\ln\frac{\mu +\sqrt{\mu^2 -g^2\sigma^2}}{g\sigma}\right].
\label{eq206'}
%\end{eqnarray}
\end{equation}
\end{widetext}
The chemical potential $\mu_c$ at which chiral symmetry
gets restored is given by the condition
$\Omega(0,\mu_c)=\Omega(\sigma,\mu_c)$.
For $\mu_c<gf_{\pi}$, the solution of (\ref{eq206}) is
$\sigma=f_{\pi}$, so (\ref{eq206}) leads to
\begin{equation}
 \mu_c=\left( \frac{\lambda\pi^2}{2} \right)^{1/4} f_{\pi}.
 \label{eq207}
\end{equation}
Since $\mu_c$ is the threshold where extended nuclear
matter forms,
a reasonable  assumption,
confirmed
by the strong-coupling QCD analysis \cite{bil2,bil3},
is
\begin{equation}
\mu_c \leq m_F        \, .
  \label{eq208}
\end{equation}
It may be easily verified that with this assumption
the pion velocity is constant and
equal to light velocity up to the transition point
$\mu_c$, where it drops to zero.

It is remarkable that Eq.\ (\ref{eq208}) yields
an upper bound on the sigma mass
\begin{equation}
 m_{\sigma} \leq \frac{2m_F^2}{\pi f_{\pi}}.
\label{eq209}
\end{equation}
If we take $f_\pi=92.4$ MeV, $m_F=340$ MeV, and saturate the bound,
we find  $m_{\sigma}= 796.5$ MeV.
In the following numerical analysis we take this
value as an input parameter.
Any other choice below this value would not alter our qualitative
picture.

Next we analyze the case of nonzero temperature.
In Fig.\ \ref{fig1} we plot
the pion velocity $v$ as a function of temperature
for various fixed $\mu$ (upper plots) and as a function
of chemical potential for various fixed $T$ (lower plots).
For each $T$ and $\mu$
the chiral condensate
 $\sigma$
is calculated by
solving  Eq.\ (\ref{eq032}) numerically.
The dashed line represents the velocity
obtained with thermodynamically unstable
solutions for $\sigma$.
    This line follows a similar backwards going line
in  the plot for $\sigma$ \cite{bil1}, which is typical of the
   first order phase transition.
The actual phase transition takes place at the
point $T_c$ (for a given $\mu=\mu_c$) or at $\mu_c$
(for a given $T=T_c$) where the two minima of the
thermodynamic potential at $\sigma=0$
and $\sigma(T_c,\mu_c)$ are leveled \cite{bil1}.
At the critical point, the pion velocity drops abruptly to zero.
We note that there is always a region of temperatures
or chemical potentials below the critical point where
the pion velocity becomes superluminal.
  The temperature $T=50$ MeV  and the chemical potential
  $\mu=160$ MeV  are chosen to represent
   a typical behavior below the critical temperature
   and the critical chemical potential, respectively.
   The value $T=3$ MeV is chosen to show the low-temperature
   behavior near the point $T=0$ which we have
   discussed above.
   If we increase the temperature, starting
   from the point ($T=0,\mu=0$) where $v=1$,
   the velocity at the beginning drops slightly below 1 and at some
   temperature $T_{\rm SL}$ of the order of 50 MeV,
   it becomes superluminal.

\begin{figure}[b]
\includegraphics{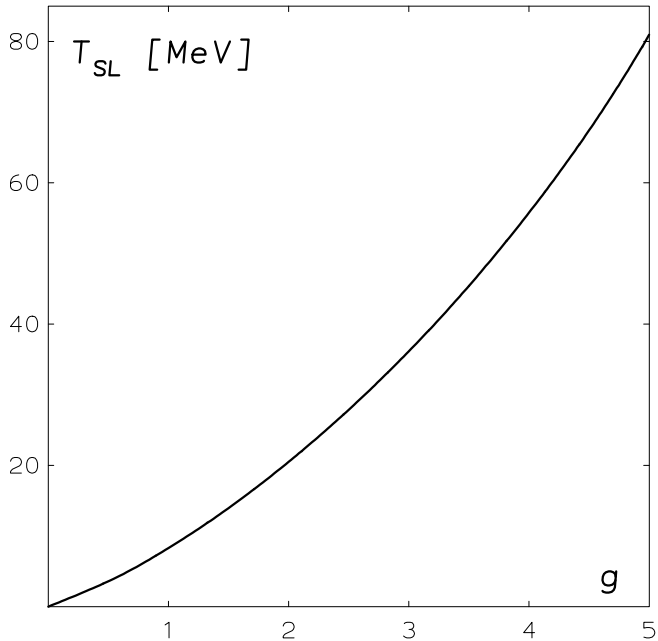}
\caption{
Superluminal onset temperature
as a function of the fermion-boson coupling $g$.
}
\label{fig2}
\end{figure}

      It is important to check how sensitive the onset of
   superluminal propagation is to the variation of the parameters
   of the model.
   In Fig.\ \ref{fig2} we plot the superluminal onset temperature
   $T_{\rm SL}$ as a function of the fermion-boson
   coupling $g$ keeping the
   boson self-coupling $\lambda$ fixed. Contrary to what one would
   naively expect, the superluminal onset shifts to larger
   temperatures with increasing $g$. The reason is that in this
temperature range the contribution of fermions (\ref{eq305}), relative
to that of bosons (\ref{eq205}), decreases with
   increasing $g$ owing to the fermion mass dependence of
   the distribution function.

\begin{figure}[t]
\includegraphics{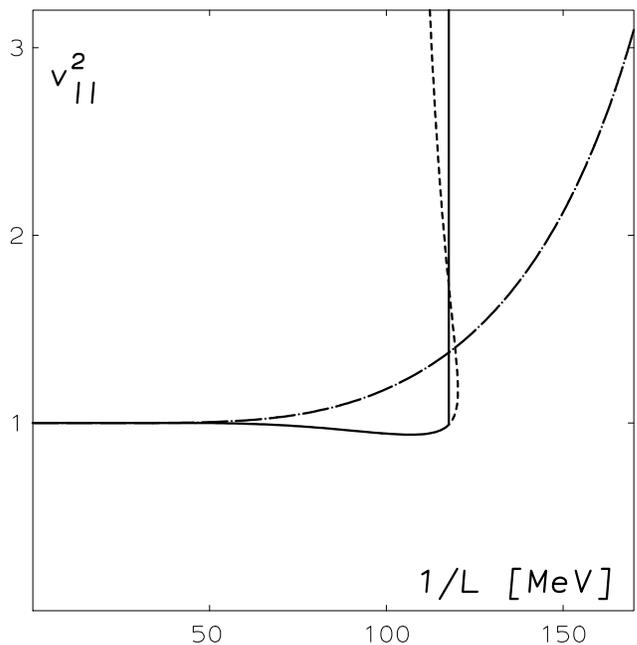}
\caption{
Pion velocity squared as a function of inverse
compactification length with
antiperiodic (solid line)
and periodic (dot-dashed line) fermions.
The dashed line corresponds to a physically
unstable solution.
}
\label{fig3}
\end{figure}

      The behavior of the pion velocity near the critical
   temperature should be analyzed with special care. It is well
   known that the phase transition in the SU(2)$\times$SU(2) linear
   sigma model at $\mu=0$ is of second order \cite{pis}
   and that the one-loop approximation
   violates this prediction by producing a weak first-order
   transition \cite{bil1,rod}.  Calculations of the
   chiral condensate in the chiral perturbation theory
   \cite{gas} show the deviation from
   nonperturbative renormalization group calculations
   \cite{ber} at temperatures of about $0.5\,T_c$.
   As our results  depicted in Fig.\ \ref{fig1}
   (upper left plot) show that
   the superluminal pion propagation starts at temperatures of
   about 50 MeV, one would perhaps be inclined to think that the
   superluminal effect was an artifact of the one-loop approximation.
   However, as we have already pointed out, the superluminal
   pion velocity is a consequence of the negative fermion loop
   contribution.
   The proportion of the fermion part
   to that of the boson depends on the parameters of the model.
   For example, a decreasing of $g$ with other free parameters
   fixed, increases the negative fermion contribution
   given by Eq.\ (\ref{eq305}), which in turn yields a superluminal
 propagation even at temperatures much below 50 MeV (Fig.\ \ref{fig2}).
   Hence, we believe that the onset of superluminal propagation
   coincides only accidentally with the window
   where  perturbation theory starts to deviate from
   nonperturbative calculations.
   For a firm judgment whether a superluminal pion
   propagation is a genuine effect or an artifact of the
   perturbation expansion one should go beyond the one-loop
   approximation or apply a nonperturbative technique
   such as Monte Carlo calculations on the lattice
   or the $1/N$ expansion
   of the O($N+1$)/O($N$) nonlinear sigma model with fermions.

Next we discuss the case
$T=0$ and $\mu=0$ with one
spatial dimension compactified
to the size
 $L\equiv\beta$.
The pion velocity $v_{||}$ parallel to the compact direction is
plotted in Fig.\ \ref{fig3} for both periodic
and antiperiodic fermions.
In the case of antiperiodic fermions,
the compactification size plays
the role of inverse temperature and chiral
symmetry gets restored at the critical size
$L_c=1/T_c$.
In this case, the velocity plot
is just the  inverse of the first plot in Fig.\ \ref{fig1}.

   For periodic fermions we find
  the minimum of the thermodynamic potential
  corresponding to  a nonzero sigma to be always below
  the minimum corresponding to $\sigma=0$.
  In this case, the right-hand side of (\ref{eq505}) is negative
  and no restoration of chiral symmetry
   takes place.
Owing to the opposite sign of the fermion contribution
the velocity $v_{||}$ is always superluminal, monotonously
approaching the velocity of light as the size of the
compactification $L$ approaches infinity.

   To gain a physical insight into why the pions become superluminal
   it is worthwhile making a comparison with the similar
   effects found for photons propagating in a medium
   and in a Casimir vacuum. In QED at either finite
   temperature or finite boundaries
   the genuine change of the speed of massless photons is
   a two loop effect related to the coherent light-by-light
   scattering \cite{bart,lat}.
   However, a one loop superluminal effect has been found
   in QED with one space dimension compactified \cite{fer}.
   Modifications of the vacuum that change the population of real and
   virtual particles introduce coherent scattering
   which decreases or increases the speed of massless photons.
   In a Casimir vacuum some of the virtual modes are eliminated
   and consequently their would be scattering \cite{lat}.
   As a result the speed of massless photons is increased.
   Roughly speaking, the negative Casimir energy density
   results in massless photons being superluminal.
   In our case, a superluminal propagation is, formally,
   a consequence of the negative fermion one-loop contribution to
  the $q_0^2$ coefficient in the inverse pion propagator (\ref{eq202}).
   It is conceivable that, for the parameters
   close to the phase transition point,
   the specific interaction between the chiral fields and fermions
   reduces the effective vacuum energy density and as a result
   the massless pions become superluminal.

\section{Summary and conclusion}
\label{concl}
We have analyzed the
 propagation of massless pions at nonzero temperature and
nonzero finite baryon density in a sigma model with two quark flavors.
By calculating the pion dispersion relation at
one-loop order we have shown
that pions in the presence of fermions
become superluminal
in a certain range of temperatures and baryon chemical
potential.

Furthermore, we have studied the case
when one of the spatial dimensions is compactified
with  fermions obeying periodic or antiperiodic boundary
conditions.
Restricting attention to $T=\mu=0$, we have calculated
the pion velocity $v_{||}$ along the compact direction.
We have found that for antiperiodic fermions,
pions will propagate superluminally or subluminally,
depending on the size of the compact dimension.
With periodic fermions, the velocity $v_{||}$
is always larger than the velocity of light.

A superluminal propagation of massless particles
may naively seem to contradict
special relativity and Lorentz invariance.
The most disturbing consequence of a superluminal propagation
would be an apparent violation of causality.
However, it has been convincingly argued
that superluminal effects realized in quantum field
theories with a nontrivial vacuum
do not lead to causal paradoxes \cite{lib}.
The basic argument goes as follows:
 once the conditions
that determine the vacuum fluctuations are fixed, the propagation
velocity in a given reference frame is {\em unique}.
This implies
that it is not possible to send signals both forwards and backwards
in time within one reference frame.
In other words, the causal loops are forbidden.

It is important to bear in mind that we have considered an idealized
situation when there is no explicit
chiral symmetry breaking of the original Lagrangian (\ref{eq100})
and pions are exactly massless in the broken
symmetry phase.
In reality, chiral symmetry is explicitly broken owing to
nonvanishing current quark masses.
   To make the pions massive we could have,
   as usual, introduced an explicit chiral symmetry
   breaking term in the original Lagrangian.
   In that case the quantity $v$, defined in
   (\ref{eq015}) and calculated in the limit $q\rightarrow 0$,
   would  no longer have the meaning of the pion velocity.
   Instead, given the dispersion relation $\omega=\sqrt{v^2q^2+m^2}$,
   with both $v$ and $m$ being  $T$ and $\mu$ dependent,
   one would define a group velocity $v_g=\partial\omega/\partial q$.
   It may easily be verified that  for  $q\ll m$
   the quantity $v_g$ is below one even if $v>1$.
   Of course, the situation could change  for
   $q\simeq m$. That would involve a calculation of $v$
   at non-zero momenta which is beyond the scope of the present
   paper.

\section*{Acknowledgments}
We thank I.\ Dadi\'c for valuable discussions.
This  work is supported  by
the Ministry of Science and Technology of the Republic of Croatia
under Contract No.\ 0098002.

\end{document}